# Magnetic reconfiguration of MnAs/GaAs(001) observed by Magnetic Force Microscopy and Resonant Soft X-ray Scattering

L. N. Coelho[1], R. Magalhães-Paniago[1,2], B.R.A. Neves[1], F.C. Vicentin[2], H. Westfahl, Jr.[2], R. M. Fernandes[2,3], F. Iikawa[3], L. Däweritz[4], C. Spezzani[5] and M. Sacchi[6].

1. Departamento de Física, Universidade Federal de Minas Gerais, CP 702, 30123-970 MG, Brazil
2. Laboratório Nacional de Luz Síncrotron, Caixa Postal 6192, 13084-971, Campinas SP, Brazil.
3. Instituto de Física Gleb Wataghin, UNICAMP, CP. 6165, 13083-970 Campinas, SP, Brazil
4. Paul-Drude-Institut für Festkörperelektronik, Hausvogteiplatz 5-7, 10117 Berlin, Germany.
5. Sincrotrone Trieste, Strada Statale 14, 34012 Basovizza, Trieste, Italy
6. Synchrotron SOLEIL, L'Orme des Merisiers, BP 48, 91192 Gis-sur-Yvette, France and Laboratoire de Chimie Physique - Matière et Rayonnement, UMR 7614, Université Pierre et Marie Curie, 11 rue Pierre et Marie Curie, 75005 Paris, France

We investigated the thermal evolution of the magnetic properties of MnAs epitaxial films grown on GaAs(001) during the coexistence of hexagonal/orthorhombic phases using polarized resonant (magnetic) soft X-ray scattering and magnetic force microscopy. The results of the diffuse satellite X-ray peaks were compared to those obtained by magnetic force microscopy and suggest a reorientation of ferromagnetic terraces as temperature rises. By measuring hysteresis loops at these peaks we show that this reorientation is common to all ferromagnetic terraces. The reorientation is explained by a simple model based on the shape anisotropy energy. Demagnetizing factors were calculated for different configurations suggested by the magnetic images. We noted that the magnetic moments flip from an in-plane mono-domain orientation at lower temperatures to a three-domain out-of-plane configuration at higher temperatures. The transition was observed when the ferromagnetic stripe width L is equal to 2.9 times the film thickness d. This is in good agreement with the expected theoretical value of L = 2.6d.

**I. Introduction**

Magnetic materials integrated to semiconductors have been intensively investigated due to their applications to spin valves and spin-injection based devices[1]. MnAs is a promising candidate for spin injection devices[2], being epitaxially compatible with GaAs based heterostructures[3,4] forming a good interface between the semiconductor and the ferromagnetic material[4]. Bulk MnAs undergoes a first order phase transition upon heating at about 40°C, from a low temperature ferromagnetic $\alpha$-MnAs to a paramagnetic $\beta$-MnAs[5]. For MnAs thin films grown epitaxially on GaAs(001) a large temperature range of phase coexistence, ~30°C, is observed, presenting the ferromagnetic $\alpha$-phase and the paramagnetic $\beta$-phase in a periodically ordered alternated groove-ridge structure[6], which can be described as a terrace-like formation. The terrace-like structure is formed to minimize the strain induced by the constraint imposed by the substrate on the MnAs film[6]. The phase coexistence has been confirmed by X-ray diffraction[6] measurements and also by atomic force microscopy (AFM)[6-9]. Several techniques have been used to investigate the magnetic properties of the MnAs films, such as magnetic force microscopy (MFM)[10-12], SQUID magnetomety[13] and ferromagnetic resonance[14]. Nonetheless, a clear understanding of the magnetic domain configuration as a function of temperature is still lacking.

Resonant magnetic soft X-ray scattering is a new technique that can combine magnetic sensitivity with structure determination[15]. By tuning the X-ray photon energy to the absorption edge of one of the atomic constituents of the sample, one can obtain direct information about the magnetic state of this atom. In this paper, we present a study of the magnetic properties of the terraces over the phase coexistence temperature range based upon magnetic force microscopy (MFM) images and polarized resonant soft X-ray scattering. Both techniques show the occurence of a magnetic moment reconfiguration, due to the change in shape anisotropy energy.

**II. Sample Characterization**

A 130nm thick MnAs sample was grown on a GaAs (001) substrate by molecular beam epitaxy under As-rich conditions at 250°C[3], with the orientation MnAs(-1100)||GaAs(001) and MnAs(0001)||GaAs(1-10). Growth details can be found elsewhere[3].

The atomic/magnetic force images were obtained with a Multimode IV microscope (Digital Instruments) using a Co-Cr covered Si probe, in Tapping Mode[16]. The magnetic measurements were performed at a distance between sample and probe of 40nm (Lift mode), where no van der Waals forces are expected to be detected. Therefore, the image is due purely to the interaction between the magnetized probe and the sample stray magnetic field only.

Figure 1 shows topographic (Fig.1a and 1c) and the corresponding magnetic (Fig.1b, and 1d) images at T=21°C and 31°C, where the difference in the magnetic profile is noticeable. Brighter areas in the topographic images indicate higher structures and are identified with the ferromagnetic terraces, about 2nm higher than the paramagnetic phase (darker stripes). A direct association can be done between the brighter areas in the topography images and the intricate pattern in the MFM images (Fig 1 b and d), showing complex domain structures in the ferromagnetic phase. Bright and dark areas in the magnetic image correspond to changes detected in the probe vibration frequency relative to its resonance frequency. Bright (dark) corresponds to an increase (decrease) in frequency, which is proportional to the derivative of the sample magnetic stray field gradient in the z direction[17].

The X-ray scattering measurements were performed at the Circular Polarization beamline of the synchrotron Elettra, Trieste, Italy. The geometry of the experiment is depicted in Fig. 2, where $\mathbf{k}_i$ and $\mathbf{k}_f$ are the incident and scattered wave vectors, respectively, with $|\mathbf{k}_f|=|\mathbf{k}_i|$. The detector was positioned at an angle $2\theta$ with respect to the incident beam and the sample at an angle $\omega$. Thus, the momentum transfer $\mathbf{q}=\mathbf{k}_f - \mathbf{k}_i$ can be decomposed into its components $q_x = 2\pi/\lambda\left(\cos(2\theta-\omega)-\cos(\omega)\right)$ and $q_z = 2\pi/\lambda\left(\sin(2\theta-\omega)+\sin(\omega)\right)$. The sample holder mounted a Peltier device for temperature control (-10 to 80°C) and an electromagnet (maximum magnetic field of 500Oe along the MnAs$(11\bar{2}0)$ direction (see Fig. 2)). The photon energy was tuned to the Manganese L$_{III}$ edge, $E = 639 eV$ $(\Delta E = 1 eV)$, to ensure that the measured signal was due mainly to the manganese atoms, with minimum influence from the substrate. In Figure 3a energy scans for two opposite directions of magnetization (denoted Mag+ and Mag-) were performed using circular polarization showing the magnetic signal at the Mn-L$_{III}$ edge. The maximum difference observed was 7% at the edge. The asymmetry ratio, defined as $(I_+ - I_-)/(I_+ + I_-)$, is shown in the right panel of Figure 3a and gives a measure of the magnetic-to-charge signal ratio. In this geometry, measurements are sensitive to the variation of the magnetization along the x-direction. Temperature, angular (Fig. 3b) and field dependent measurements were performed at 640eV, where a maximum magnetic contrast is observed.

### III. Results and Analysis

#### A. *Magnetic Force Microscopy*

On the topographic image (Figs. 1a and 1c), one notices that the bright stripes narrow as the temperature rises from 21$^{o}$C to 31$^{o}$C, an indication that the *α*-MnAs phase terrace shrinks. In Fig. 1b, for $T = 21^{o}C$, the predominant magnetic domain configuration is a meander-like structure with alternating bright and dark areas, referred to as type I, as in Ref 10. In Fig. 1d, this structure gives way to a line-shaped one, with linear structures along the terraces – direction [0001] of MnAs, referred to as type II. The change of the feature in the magnetic profile as the temperature increases follows the decrease of the ferromagnetic terrace width, as shown in the topographic images. The rearrangement of magnetic structure in MnAs films has been observed in previous works and it is associated to the change of the terraces width[10]. The schematic drawing of Fig. 4 is based on that work and relates each magnetic configuration to the magnetic image observed by MFM. Such structures have been explained as follows: at temperatures where the ferromagnetic terraces are wide enough, shape anisotropy favors the in-plane alignment of the magnetic moments (type I). As the terraces become narrower, the out-of-plane orientation of magnetic moments is energetically favorable (type II), forming long stripes of two and three domains along the y-direction. This accounts for the noticeable change in the magnetic images. In order to reassure the magnetic configuration dependence on terrace width, a rather wide terrace has been marked with an arrow in Fig. 1d, where all other terraces have line-shaped structures except for this one. Such magnetic reorientation has been suggested earlier[10] but no confirmation by an alternative experimental technique has been reported so far.

### B. Resonant soft X-ray scattering

Primarily, $q_x$-scans (rocking scans) were performed to detect long range structural correlations between the terraces. These correlations produce satellite peaks on both sides of the specular reflection (see Fig. 5) at $q_x = 2\pi/s$, where $s$ is the modulation period of the terrace structure. Following Holy *et al.*[18], the intensities of these peaks can be written as a function of the terrace structure as $I(q_x, q_z) \propto |D(q_x)|^2 |F(q_x, q_z)|^2$, where $F(q_x, q_z)$ is the Fourier transform of the terrace height profile and $D(q_x)$ is the correlation function of different sets of two terraces averaged over the whole sample[19]. The scattered intensity can be written as a function of the terrace period $s$ and of the width $L$ of the $\alpha$-phase[19]:

$$I(q_x, q_z) \propto \frac{L^2 + (s-L)^2 + 2\cos(q_z h) L(s-L)}{q_x^2 + \sigma^2} + 16\sin^2\left(\frac{q_z h}{2}\right) \sum_{n=-2, n\neq 0}^{n=2} \frac{\sin\left(\pi n L/s\right)}{|n|^3 \left(2\pi/s\right)^2} \frac{}{\left(\frac{q_x}{n} - \frac{2\pi}{s}\right)^2 + \sigma^2} \quad (1)$$

where $h$ is the terrace height (about 2.0nm from AFM images and constant with temperature) and $\sigma$ is the peak width and corresponds to the inverse of the correlation length. By fitting $q_x$-scans, it is possible to determine the temperature dependent values of $L$ to be used for fitting the hysteresis loops (see below). Values of $s = 631 nm$ and $\sigma = (1.5 \pm 0.2) \times 10^{-3} nm^{-1}$ were found not to depend on temperature.

The intensity of the satellite peaks is also sensitive to the magnetization state of the terraces. At the satellite peak, only periodic arrangements contribute to the scattering, hence both the charge and the magnetic configurations must be replaced by their Fourier transforms. Following Lee *et al.*[20], the scattered intensity is given by:

$$I(\mathbf{q}) \propto \left| \sum_\alpha e^*_{s\alpha} e_{i\alpha} \rho_{eff}(\mathbf{q}) - i \sum_{\alpha\beta\gamma} e^*_{s\alpha} e_{i\beta} \epsilon_{\alpha\beta\gamma} B\, M^{(1)}_\gamma(\mathbf{q}) + \sum_{\alpha\beta} e^*_{s\alpha} e_{i\beta} C\, M^{(2)}_{\alpha\beta}(\mathbf{q}) \right|^2 \quad (2),$$

where $\alpha, \beta, \gamma$ are Cartesian components, $e_{i\alpha}$ and $e_{s\alpha}$ are the $\alpha$-th component of the polarization vectors of the incident and scattered beams, respectively, and $\epsilon_{\alpha\beta\gamma}$ is the antisymmetric Levi-Civita symbol. $\rho_{eff}(\mathbf{q})$ is the charge Fourier transform of $\rho_{eff}(\mathbf{r})$, responsible for the satellite peaks, which can be separated into two parts $\rho_{eff}(\mathbf{r}) = \rho_0(\mathbf{r}) - r_0^{-1} A n_m(\mathbf{r})$, one coming from non-resonant atoms $\rho_0(\mathbf{r})$ and one containing the resonant part proportional to the density of magnetic atoms $n_m(\mathbf{r})$, where $r_0$ is the classical Thomson scattering length. The factor $A = f_0 + 3\lambda/8\pi (F_{1+1} + F_{1-1})$ contains the resonant component of the $F_{L\Delta m}$ factors, which are dipolar transition probabilities determined by the Fermi's golden rule[20]. $f_0$ is the usual Thomson charge scattering. Factors $B = 3\lambda/8\pi (F_{1+1} - F_{1-1})$ and $C = 3\lambda/8\pi (2F_{10} - F_{1+1} - F_{1-1})$ are responsible for the resonant effect in the magnetic terms of the scattering amplitude. Magnetic contributions to the scattering amplitude can be written as a function of the magnetic profile $M(\mathbf{r})$ of the sample, in a linear term and a quadratic term:

$$M_\gamma^{(1)}(\mathbf{q}) = \int d\mathbf{r} e^{-i\mathbf{q}\cdot\mathbf{r}} n_m(\mathbf{r}) M_\gamma(\mathbf{r}) \tag{3.b}$$

$$M_{\alpha\beta}^{(2)}(\mathbf{q}) = \int d\mathbf{r} e^{-i\mathbf{q}\cdot\mathbf{r}} n_m(\mathbf{r}) M_\alpha(\mathbf{r}) M_\beta(\mathbf{r}) \tag{3.b}$$

The Fourier transform in these equations shows the relationship between the scattered intensity and the magnetic profile. In contrast to magnetometry techniques, such as SQUID or Kerr effect measurements, which give the average magnetization, the magnetic signal of the satellite peak is only related to the periodic arrangement of the magnetic domains of the terraces. The magnetic measurements on the satellite peak show that the reorientation observed is common to all ferromagnetic terraces and only cases where the magnetic periodicity is equal to the structural periodicity are measured. Considering the experimental conditions $C \ll B$ [20] and the contribution of $M^{(2)}(\mathbf{q})$ is negligible. Therefore the magnetic contribution to the X-ray scattered intensity is proportional to $M_x^{(1)}(q_x = 2\pi/s) = M_x^{(1)}(9.96\times10^{-3} nm^{-1})$, which corresponds to the mean periodic magnetization of the terraces in the x direction. Figure 3b shows rocking scans with the sample before (demagnetized) and after (ordered) applying an external magnetic field, at 15°C. Comparing the two rocking scans, we observed the difference in intensity of the diffuse satellite peaks as the magnetic state is changed, thus indicating the effects of the periodic magnetic structure on the X-ray scattering signal.

We measured the intensity of the first satellite peak as a function of an external magnetic field applied along the x-direction, from -300Oe to 300Oe. The magnetic field range used is enough to achieve the saturation of the magnetization, where the coercive field is less than 100Oe. The X-ray magnetic signal is sensitive to the direction of the magnetization therefore we obtain different intensity for positive and negative magnetic field, resulting in a hysteresis loop, as shown in Fig. 5. The hysteresis measurements were performed with the detector at 11° and sample at 4.56° for different temperatures and a rocking scan was also performed for each temperature. Figure 5 shows hysteresis loops and rocking scans measured as a function of temperature. At low temperatures, below 23°C, the hysteresis loop is square-shaped, while at 25°C, the it changes to s-shaped. At 33°C, the hysteresis loop reduces its amplitude till it vanishes, indicating that the ferromagnetic phase is transforming progressively into the paramagnetic $\beta$-phase. The satellite intensity dependence on temperature is addressed elsewhere[19].

The evolution of the feature of the hysteresis loops versus temperature can be associated to the change in the magnetic moment configuration observed in the MFM images. The change of the magnetic configuration is related to the shape anisotropy energy[21] due to the change of the width of the ferromagnetic terrace. The MnAs crystalline anisotropy defines the y-direction as a uniaxial hard axis for the magnetization, forcing the magnetic moment to lie in the xz-plane. On the other hand, the values for the crystalline anisotropy constants of the x and z directions of MnAs are nearly the same[14] and the preferred magnetization direction will be determined only by the shape anisotropy energy. For a MnAs thin film in the $\alpha$-phase, the magnetic moments tend to lie in the film plane and in the x direction. When the paramagnetic phase starts to appear and the terraces are formed, one has to consider the $\alpha$-MnAs stripes as slabs such as the one depicted in Fig. 6. The favored magnetic axis depends thus on the ratio p between the slab width L and thickness d of the ferromagnetic terrace (i.e., p = L/d). Once the terraces are narrow enough, the shape anisotropy favors magnetic moments aligned perpendicular to the film plane, in the z direction. This suggests a change in the preferred axis of magnetization as temperature rises.

In order to determine which magnetic configuration of the MnAs stripes should be compatible with the experimental results, one must determine the shape anisotropy energy as a function of the terrace width L. The shape anisotropy energy was calculated considering the stripe an infinite slab in the y-direction (hard axis) with width L (x-direction) and thickness d (z-direction) for a number of N domains. The demagnetization factors $D^{(N)}(p)$ in the x and z directions for a number of N domains were found modeling the magnetization of the domains as a square-wave[22]. More elaborate magnetic models can be found in Ref. 23. According to the schematic diagram shown in Fig. 6b, the in-plane one domain configuration will be referred to as IP1 and the out-of-plane N-domain configuration as OPN. It is possible to identify IP1 with the suggested configuration drawn in Fig. 3a (meanders – type I), and both OP2 and OP3 with Fig. 3b (linear structures – type II). The expression for the demagnetizing factor is:

$$D_{zz}^{(N)}(p) = \frac{4}{\pi} p \int_0^\infty \tan^2\left(\frac{\xi}{2p}\right)\left[1-(-1)^N \cos(\xi)\right] \sinh\left(\frac{\xi}{2p}\right) d\xi \qquad (4)$$

where $p = L/d$ (see Fig. 7). The demagnetizing factor $D_{xx}^{(1)}(p)$ in the x direction can be obtained from Eq. 5 by replacing $p$ by $1/p$. The shape anisotropy energy per unit length was then calculated for different magnetic configurations as shown in Fig. 7. Apparently, the larger the number of domains along the z-direction the lower the energy. However, no more than 3 domains are observed in the MFM images, suggesting that the reconfiguration involves a transition from the IP1 configuration to OP3. This is probably due to the energy cost of formation of more domain boundaries. It is noticeable that the expected transition from IP1 to OP3 is predicted for p=2.6, which is close to the observed transition at p=2.9, determined from the fit of the rocking scan shown in Fig. 8.

Based upon this information, the hysteresis loops were fitted using the Stoner-Wohlfahrt model[24,25] considering the energy competition between IP1 and OP3:

$$E(p,\varphi,H) = \left[\frac{1}{2}M^2 4\pi \left(D_{zz}^{(3)}(p) - D_{xx}^{(1)}(p)\right)\right]\sin^2\varphi - HM\cos\varphi \qquad (5)$$

where $\varphi$ is the angle between the magnetization and the x-direction, M is the saturation magnetization (taken to be 0.67MA/m [21], corresponding to 2.5$\mu_B$/Mn atom) and H is the applied magnetic field (see Fig. 2). The first term in Eq. 6 can be associated with a uniaxial anisotropy energy term that changes the easy axis direction with temperature. At low temperatures, the difference $D_{zz}^{(3)} - D_{xx}^{(1)}$ is negative and the preferred axis is in the x direction. As temperature rises, the ratio *p* decreases, causing the anisotropy easy axis to flip from the x direction to the z direction (see Fig. 7). Based on the change of anisotropy axis direction, we have fitted the hysteresis curves with a Gaussian distribution for p around $p = L_c/d$, where $L_c$ is the best fit parameter of the rocking scan using Eq. (1), as shown in Fig. 8 (left), along with hysteresis loops fitted according to this model (right). The Stoner-Wohfahrt model sets a limiting value for the coercive field given by $4\pi \left( D_{zz}^{(3)}(p) - D_{xx}^{(1)}(p) \right) M$ [25], which is much higher than observed. Therefore, the magnetic field in all scans was rescaled to fit the observed coercive field.

The squared-shape loop at 23ºC was fitted using a terrace width distribution of 0.05$L_c$. Since the terraces are much wider than thicker, the size distribution plays a secondary role, with all terraces with the easy axis of magnetization in the x-direction. At 25ºC, the size distribution becomes important, with some terraces presenting the easy axis in the x-direction and others in the z-direction. Associating the shape anisotropy energy with an internal demagnetizing field, the sample may be compared with an ensemble of randomly oriented particles with uniaxial anisotropy. The demagnetizing fields associated with the terraces add vectorialy and the corresponding effect is that of terraces with easy axes that span the whole xz-plane. As a result, one obtains the s-shaped hysteresis loop.

**IV. Conclusions**

We used polarized resonant soft X-ray scattering and magnetic force microscopy techniques to investigate the magnetic and structural properties and their evolution with the temperature in MnAs thin films during the phase coexistence temperature range. Each method provides particular and complementary information, making them excellent tools to investigate magnetic thin films. Magnetic force microscopy profiles show meander-like magnetic structures (in-plane domains) that change to line-shaped structures associated with out-of-plane domains as temperature increases. This reconfiguration is confirmed by magnetic field and temperature dependent X-ray resonant scattering with a distinct change in the shape of the hysteresis loops at 25°C and is explained by a simple model based on magnetic shape anisotropy energy. The Magnetic Force Microscopy images suggest a transition from an in-plane one domain configuration to a 3 domains out-of-plane one. Demagnetizing factors calculated for different magnetic configurations are used to fit the hysteresis loops obtained by X-ray resonant scattering, yielding a transition when p=2.9. The X-ray measurements also show that these configurations of the magnetic domains present long range periodicity.

**Acknowledgments**

The authors thank M. Kästner and C. Herrmann (Paul Drude Institute) for sample growth. C. Grazioli and S. Rinaldi (Circular Polarisation Beamline) and the personnel of Sincrotrone Trieste helped us during the X-ray experiments. This research was supported by CNPq, FAPESP, FAPEMIG and a bilateral CNPq-CNRS agreement.

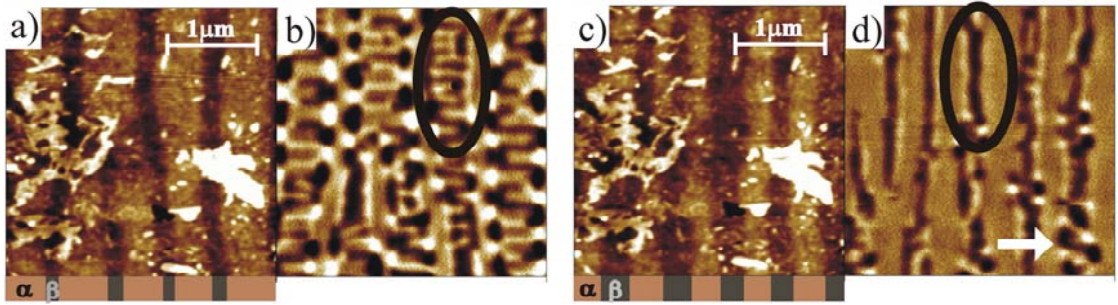

Figure 1 – (Color online) Two 3 x 3μm² dual scans of the sample (130nm) done in lift mode at different temperatures, 21°C (left) and 31°C (right). It is possible to correlate the brighter areas in the topography images (a and c) with the intricate pattern in the magnetic force images (b and d). The ferromagnetic α-phase and the paramagnetic β-phase are indicated bellow the topographic image. Comparing the magnetic images, the meander like structure at low temperature disappears at high temperature, as in the terrace indicated by the ellipse. The arrow on the lower right side (d) indicates a wider terrace, where the magnetic reorientation has not yet occurred.

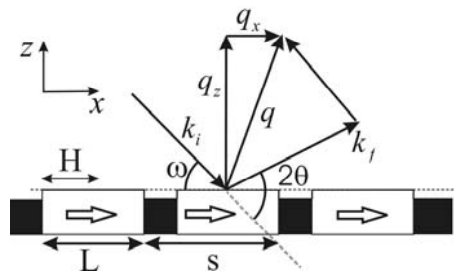

Figure 2 – (a) Schematic of the X-ray scattering experiment. H is the direction of applied field H. L is the width of the ferromagnetic terrace and s the period of the terrace. The dark area represents the paramagnetic β-phase. Light is circularly polarized and sensitive to magnetic moments in the x-direction, indicated by arrows.

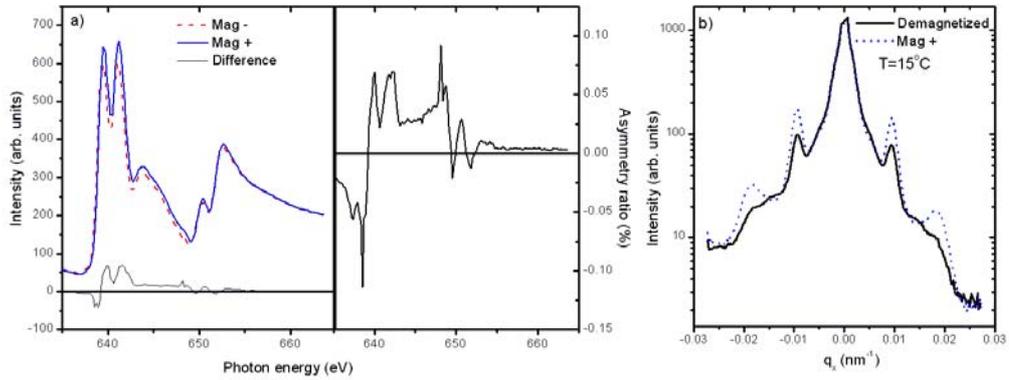

Figure 3 – (Color online) a) Energy scan (left) at the satellite peak angular position, across the 2p-edge of Mn for two opposite magnetization directions and (right) corresponding asymmetry ratio (difference divided by the sum). b) Rocking scans (specular intensity removed) at T=15°C for two different magnetic states, the sample demagnetized (full line) and after a pulse of 290Oe parallel to the sample plane (dashed).

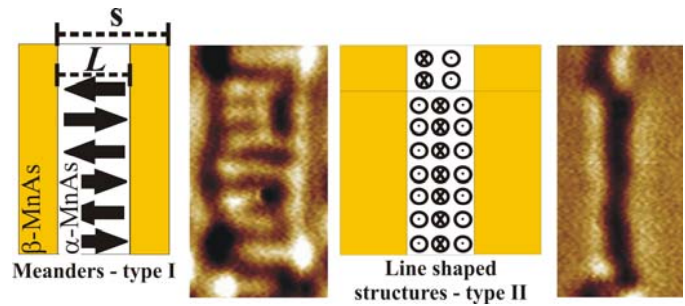

Figure 4 – (Color online) Schematic representation of the magnetic configurations of α-MnAs at low temperature (Type I) and high temperature (Type II). These configurations would each result in the magnetic images shown to their right, zoom images from the marked areas in Fig. 1b and 1d.

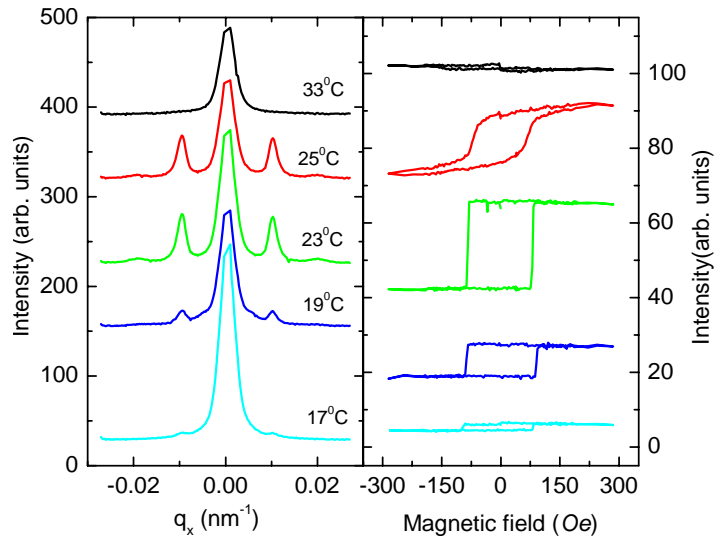

Figure 5 – (Color online) Rocking scans for different temperatures and the respective hysteresis loops. One sees the change in shape of the hysteresis curves from square in T=23°C to s-shaped in T=25°C.

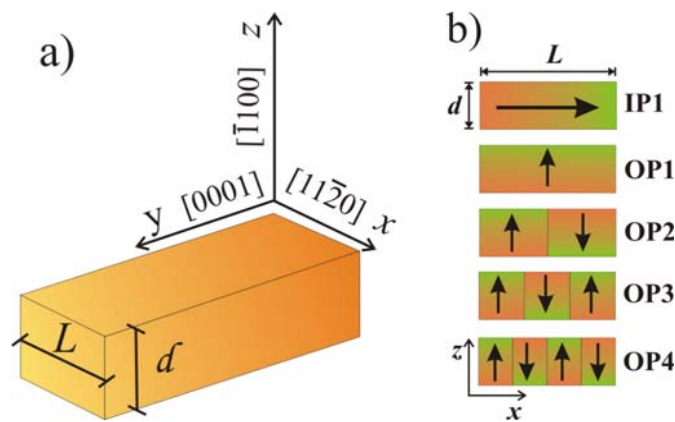

Figure 6 – (Color online) a) Slab representing the ferromagnetic terrace during the coexistence phase, with variable width $L$ in the x-direction (11-20) and constant thickness d in the z-direction (-1100). The y-direction (0001) is a hard magnetic axis and is not considered for shape anisotropy energy configurations. b) Representations of the magnetic configurations.

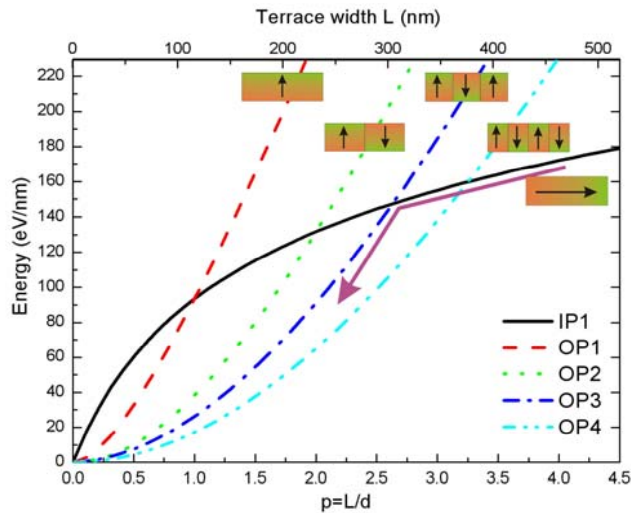

Figure 7 – (Color online) Shape anisotropy energy per unit length as a function of the ratio *p* for a 130nm thick sample. For large *p*, the in-plane moment configuration has lowest energy. When the terrace width is 2.6 times its thickness, the 3-domain out-of-lane configuration prevales. Based on MFM analysis, configurations with more than 3 domains are not expected to occur.

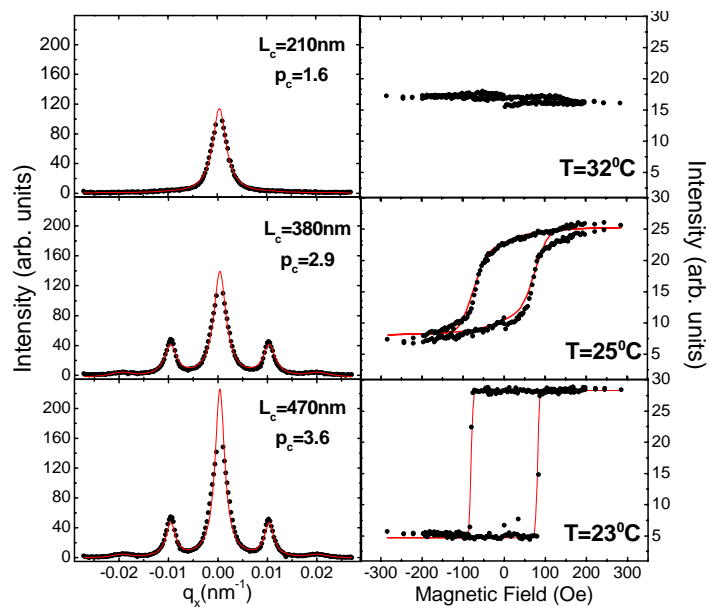

Figure 8 – (Color online) Temperature dependent rocking scans (left) and corresponding hysteresis loops (right). The fitted width $L_c$ of the terrace was obtained from the rocking scan and used as the center of the Gaussian size distribution in the hysteresis fit ($p_c = L_c/d$).

All hysteresis loops were shifted and centered at the same position and the magnetic field of the fit was rescaled to fit the loops.